\documentclass[12pt]{iopart}
\usepackage{iopams,graphicx}
\begin{document}

\title[Harmonic Ampitudes From Grid Data]{Spherical
Harmonic Amplitudes From Grid Data}

\author{Mark E. Rupright}

\address{Florida Atlantic University\\
The Harriet L. Wilkes Honors College \\
5353 Parkside Dr. Jupiter, FL 33458}

\ead{rupright@fau.edu}
\begin{abstract}
The problem of resolving spherical harmonic components from
numerical data defined on a rectangular grid has many applications,
particularly for the problem of gravitational radiation extraction.
A novel method due to Misner improves on traditional techniques by
avoiding the need to cover the sphere with a coordinate system
appropriate to the grid geometry. This paper will discuss Misner's
method and suggest how it can be improved by exploiting local
regression techniques.
\end{abstract}

\maketitle

\section{Introduction}

Spherical harmonic decomposition is an important tool in
computational science. In numerical astrophysics, for example,
harmonic decomposition can serve as the basis of an algorithm for
extracting gravitational waveforms from simulations of radiating
systems.~\cite{POBM,Fiskeetal,LSU}

Harmonic decomposition of data defined on a two-sphere is a
straightforward computation. Consider a scalar field
$\Phi(t,r_0,\theta,\phi)$ defined on a two-sphere of radius $r_0$.
The field can be expanded in terms of scalar spherical harmonics as
\[ \Phi(t,r_0,\theta,\phi) = \sum_{\ell,m} \Phi^{\ell\,m}(t,r_0) \,
Y_{\ell\,m}(\theta,\phi) \; . \] The orthonormality of the spherical
harmonics with respect to integration over the sphere can be used to
compute the harmonic amplitudes
\begin{equation}
\Phi^{\ell\,m}(t,r_0) = \oint \bar{Y}_{\ell\,m}(\theta,\phi) \,
\Phi(t,r_0,\theta,\phi) \, \rmd^2 \Omega\; ,
\label{Eq:HarmonicAmplitudes}
\end{equation}
where the bar denotes complex conjugation.

Of interest for numerical analysis are cases in which the field
$\Phi$ is not known as an analytic function on the sphere, but
rather as the solution of, for example, a wave equation evolved on a
discrete set of points $\{ x_i \}$ which comprise a
three-dimensional grid for the numerical simulation. In general
these grid points will not lie on the sphere $r=r_0$. Therefore,
computing the harmonic coefficients $\Phi^{\ell\,m}$
from~(\ref{Eq:HarmonicAmplitudes}) requires a method of determining
field values on the integration sphere.

The most obvious method of computing $\Phi^{\ell\,m}$ from grid data
is first to interpolate data from the grid onto coordinate patches
covering the two-sphere, then compute a numerical integral of the
interpolated data against the conjugate spherical harmonics. Because
the grid data is the source of all information about $\Phi$, it is
important that the coordinate patches on the two-sphere adequately
represent the distribution of three-dimensional grid points near the
sphere. For example, a spherical coordinate grid discretized in $\{
\theta,\phi \}$ will over-represent points near the poles and
under-represent points near the equator.

Previous investigations of the the gravitational radiation
extraction problem used a pair of stereographic coordinate patches
to cover the sphere~\cite{POBM}. These patches, due to G\'omez,
\etal~\cite{Pitt}, overlapped at the equator and special care was
required to compute the integrals of~(\ref{Eq:HarmonicAmplitudes})
correctly. A useful alternative is to use ``cubed sphere''
coordinates like those used by Zink, \etal~\cite{LSU}. In these
coordinates the six patches represent cartesian grids, deformed to
cover a sphere, which join nicely at the patch boundaries. The
distribution of points in these patches is approximately uniform and
close to the distribution of nearby grid points.

Misner has proposed an attractive alternative approach to the
multipole decomposition problem that replaces the traditional
two-step interpolation/integration method with a single volume
integration step for each mode~\cite{Misner04}. In addition to the
computational simplification offered by this approach, Misner's
method completely bypasses the need to choose a coordinate grid
covering the two-sphere. Instead, it relies directly on values of
$\Phi$ on the grid points surrounding the sphere. The relative
weight of each point in the calculation depends only on its
proximity to the sphere.

Fiske~\cite{Fiske05} has recently presented an analysis of symmetry
and convergence properties of Misner's method, and Fiske,
\etal~\cite{Fiskeetal} have demonstrated its usefulness in
extracting gravitational radiation from cubical grids with fixed
mesh refinement.

This paper presents an alternative analysis of Misner's algorithm
which exploits the fact that it can be cast as a weighted local
least-squares regression procedure. This approach somewhat
simplifies the computational aspects of Misner's method, and
improves the accuracy of the calculations.

While the analysis in this paper will assume complex-valued
harmonics, test computations on a real-valued scalar field will be
performed using real-valued spherical harmonics for simplicity. The
test function will be the analytical quadrupole solution of a
three-dimensional linear scalar wave equation. Specifically, the
test function will be $\Phi(t,r,\theta,\phi) = \Phi^{2 0}(t,r) Y_{2
0}(\theta,\phi)$, where $\Phi^{2 0}$ is constructed from a
generating function of the form $(x/\lambda^2)
\exp(-x^2/\lambda^2)$, and $x = t \pm r$. The initial data is
time-symmetric and we choose unit wavelength ($\lambda = 1$) and
amplitude. This is similar to the wave evolved by Fiske,
\etal~\cite{Fiskeetal}, but here $\Phi$ is the solution of a linear,
scalar wave equation. The computational grid extends from $-8$ to
$8$ with uniform spacing $h$ in each dimension. Multipole amplitudes
are computed from this test data on a two-sphere of radius $r_0 =
6$. Most computations are performed in the first octant with even
reflection symmetry imposed at the coordinate plane boundaries.

\section{Misner's Algorithm}
\label{Sec:Misner}

\subsection{Continuum Limit}

Recognizing that a surface integral
like~(\ref{Eq:HarmonicAmplitudes}) can be treated as the derivative
of a volume integral, Misner approaches the problem of computing
harmonic coefficients by integrating data over a specified volume
surrounding the two-sphere. The volume of interest is a spherical
shell, $S$, of half-width $\Delta$ that surrounds the sphere: \[ S =
\left\{ x \equiv \{r,\theta,\phi\}\, | \; r \in \left[ r_0 - \Delta,
r_0 + \Delta \right] \right\}\; .\]

He then expands the field $\Phi$ not only in terms of orthonormal
basis functions in $\{\theta,\phi \}$ (spherical harmonics), but
also in terms of radial basis functions $R_n(r)$ ($n=0, \ldots,
\infty$) that are orthonormal with respect to integration over the
radial interval $[r_0-\Delta, r_0+\Delta ]$:
\begin{equation}
\int_{r_0-\Delta}^{r_0+\Delta} \bar{R}_n(r)\, R_m(r) \, r^2 \rmd r =
\delta_{n m} \; .\label{Eq:ROrthonormal}
\end{equation}
The tensor product of the radial and angular basis functions
represents a new set of three-dimensional basis functions that are
orthonormal with respect to integration over the shell $S$. These
can be written as
\[ Y_A \, \equiv \, Y_{n \, \ell \, m} = R_n(r) \, Y_{\ell \, m}
\left( \theta, \phi \right) \;,\] where the single index $A \equiv
\left\{ n \, \ell \, m \right\}$ denotes the three indices that
identify the three-dimensional basis functions.

Specifically, Misner uses Legendre polynomials $P_n$ to generate the
(real-valued) radial basis functions
\[ R_n(r) = \frac{1}{r}\, \sqrt{\frac{2 n + 1}{2 \Delta}} \,
P_n\left(\frac{r-r_0}{\Delta}\right) \;, \] which
satisfy~(\ref{Eq:ROrthonormal}). The choice of Legendre polynomial
as the basis for $R_n$ is not unique and other choices will be
investigated later.

The integral over $S$ defines an inner product for functions defined
on the shell:
\[ \left< f | g \right> \, \equiv \, \int_S \bar{f}(x) \, g (x) \,
\rmd^3 x \; ,\] where the bar denotes complex conjugation. The
orthonormality relation can be expressed as $\left< Y_{A^\prime} |
Y_A \right> \, = \, \delta_{A^\prime \, A} \, \equiv \,
\delta_{n^\prime n} \, \delta_{\ell^\prime  \ell} \,
\delta_{m^\prime  m}$.

The expansion of $\Phi$ in terms of these basis functions is
\[ \Phi(t,x) = \sum_A \Phi^A(t) \, Y_A(r,\theta,\phi)
\; ,\] where the coefficients of the expansion are $\Phi^A(t) =
\left< Y_A | \Phi \right>$. The harmonic coefficients at $r_0$ are
expressed in terms of the coefficients $\Phi^A$ as:
\begin{equation}
\Phi^{\ell\,m} (t,r_0) = \sum_n R_n (r_0) \, \Phi^{n\,\ell\,m} (t) =
\sum_n R_n(r_0) \, \left< Y_{n\,\ell\,m} | \Phi \right>\; .
\label{Eq:HarmonicCoefficients}
\end{equation}

It is useful to re-cast~(\ref{Eq:HarmonicCoefficients}) in terms of
a set of projection functions,
\begin{equation}
p^{\ell\,m} (x) = \left\{ \sum_n R_n(r_0) R_n(r) \right\} {\bar
Y}_{\ell\,m} (x) \; , \label{Eq:ProjectionFunctions}
\end{equation}
that act on $\Phi$ under the inner product to give the harmonic
amplitudes:
\begin{equation}
\Phi^{\ell\,m} (t,r_0) = \left< p^{\ell\,m} | \Phi \right> = \int_S
\, p^{\ell\,m}(x) \, \Phi(t,x) \, \rmd^3 x  \; .
\label{Eq:ProjectionOperation}
\end{equation}

\subsection{Numerical Approximation}

Misner's algorithm is based on a specific method of converting the
continuum integral above into a numerical integral over the grid
points located within the shell $S$. By assuming a cell-centered
grid of uniform spacing $h$ (cell volume $h^3$), one can approximate
a volume integral by a sum that is equivalent to a ``midpoint''
quadrature:
\begin{equation}
\sum_{x_i \in S} \bar{f} (x_i) \, g (x_i) \, h^3 \, \approx \,
\int_S \bar{f}(x) \, g(x) \, \rmd^3 x \; ,
\label{Eq:NumericalInnerProduct}
\end{equation}
ignoring effects near the boundary of $S$.

Of course some points near the shell boundary belong to cells that
do not lie entirely within $S$, while other points located just
outside the boundary belong to cells that lie partially within the
shell. To account for this effect, Misner adds to the
sum~(\ref{Eq:NumericalInnerProduct}) all points a distance less than
$\frac{1}{2} h$ {\it outside\/} of S. The correct inner product uses
points from a larger shell, which we shall denote $S_+$, defined to
be the set of all grid points that lie within a radial distance
$\delta \equiv \Delta + \frac{1}{2} h$ of $r_0$. Furthermore, all
points that lie ``near'' the shell boundary are weighted based on
the partial volume of their cell that lies within $S$.

Keeping track of these partial volumes is complicated, but a simple
approximation is to treat each point near the boundary as if it lay
on a coordinate axis. This makes the partial volume a linear
function of r:
\[
w(r_i) = \left\{ \begin{array}{l l l}
0 & & | r_i - r_0 | > \delta \\
h^3 & & | r_i - r_0 | < \delta - h \\
\left( \delta - | r_i - r_0 | \right) h^2 & & \mathrm{otherwise,}
\end{array} \right.
\]
where $r_i$ is the radial coordinate of point $x_i$. Other choices
of weight function will be discussed below.

In Misner's method, computations of the harmonic coefficients from
grid data are based on the numerical inner product
\begin{equation}
\left< f | g \right> \, \equiv \, \sum_{x_i \in S_+} \bar{f} (x_i)
\, g (x_i) \, w(r_i) \; . \label{Eq:WeightedInnerProduct}
\end{equation}

In the continuum limit, the inner product of basis functions is
equivalent to the identity matrix. However, because the weighted sum
of~(\ref{Eq:WeightedInnerProduct}) is only an approximation to the
continuum integral, the basis functions $Y_A$ are not strictly
orthonormal and the numerical inner product defines a matrix $G_{A
B} = \left< Y_A | Y_B \right>$ that approximates the identity. While
some matrix elements of $G_{A B}$ will be identically zero due to
symmetries, the remaining elements will only approximate their
continuum values to a degree determined by the numerical
approximation. As a result, using the numerical inner
product~(\ref{Eq:WeightedInnerProduct}) in the calculation
of~(\ref{Eq:HarmonicCoefficients}) gives a convergent set of values
of $\Phi^{\ell\,m}(t,r_0)$, but could result in mode mixing.

We can avoid this by using the inverse of the inner product matrix,
$G^{A B} \equiv \left(G_{A B} \right)^{-1}$, to define a set of {\it
dual\/} basis functions $Y^A \,=\, \sum_B G^{A B} \, Y_B$ that are
orthonormal to the original basis with respect to the inner product:
\[ \left< Y^A | Y_B \right> \,=\, \sum_C G^{A C}
\, \left< Y_C | Y_B \right> \,=\, \sum_C G^{A C} G_{C B} \,=\,
\delta^A_B \; . \]

Using the dual basis functions, define the expansion coefficients
$\Phi^A$ using the inner product with the dual basis
\begin{equation}
\Phi^A \,\equiv\, \left< Y^A | \Phi \right> \,=\, \sum_B G^{A B}
\left< Y_B | \Phi \right> \;, \label{Eq:AdjointCoefficients}
\end{equation}
using the numerical inner product~(\ref{Eq:WeightedInnerProduct}).
This choice defines an approximation to the original data, $\hat\Phi
\,\equiv\, \sum_A \Phi^A Y_A \,\approx\, \Phi$. For a given {\it
finite\/} collection of basis functions, the approximation
$\hat\Phi$ minimizes the (weighted) sum of the squared differences
between $\Phi(x_i)$ and $\hat\Phi(x_i)$ over the grid points in
$S_+$. In other words, Misner has replaced the
interpolation/spherical integration with a least squares procedure
to compute spherical harmonic coefficients.

The projection functions of~(\ref{Eq:ProjectionFunctions}) can be
computed at the beginning of a simulation from the dual basis
functions $Y^A$ and stored for future use. At each time step of a
simulation, the approximate spherical harmonic amplitudes
$\Phi^{\ell\,m}(t,r_0)$ are computed from the numerical inner
product of the projection functions with the field $\Phi$
\[ \Phi^{\ell\,m}(t,r_0) = \left< p^{\ell\,m} | \Phi \right> =
\sum_{x_i \in S_+} p^{\ell\,m}(x_i) \, \Phi(t,x_i) \, w(r_i) \; .\]

\section{The Least-squares Perspective}
\label{Sec:LeastSquares}

In their analyses, Misner and Fiske focus moslty on the relation
between the weighted sums and the underlying three-dimensional
integrals they approximate. It is useful, however, to consider
Misner's method strictly as a least-squares problem.

\subsection{Matrix Form}
Focusing on the least-squares aspect of the method makes it easier
to see how to compute the projection operators that will be used in
the computation of the harmonic amplitudes. It is useful to re-cast
the algorithm in the traditional matrix notation of linear
regression analysis.

Let $N$ be the number of points in the computational shell $S_+$,
and $M$ be the number of independent modes $A \equiv \{n, \ell, m\}$
over which the three-dimensional harmonic expansion is performed.
Let $\bi{\Phi}$ be the $N \times 1$ column vector of values of
$\Phi$ at all points in $S_+$. Similarly, let $\bi{Y}$ be the $N
\times M$ design matrix, where each column of $\bi{Y}$ is one of the
$M$ basis functions $Y_A$ evaluated at all points in the shell.
Finally, let $\bi{W}$ be the $N \times N$ diagonal matrix of weights
$w_i$. Then $G_{A B}$ is easily computed as $\bi{G} \equiv
\bi{Y}^\dag \bi{W} \bi{Y}$, where $\bi{Y}^\dag$ is the Hermitian
conjugate of $\bi{Y}$.

Denote the set of expansion coefficients $\Phi^A$ as the ($M \times
1$) column vector $\bi{F}$. The vector of approximate values,
$\hat\Phi \approx \Phi$ in $S_+$, is $\hat\bi{\Phi} = \bi{Y}
\bi{F}$. The weighted sum of squared residuals (SSR) between $\Phi$
and $\hat\Phi$ is $\left( \bi{\Phi} - \bi{Y} \bi{F} \right)^\dag
\bi{W}\, \left( \bi{\Phi} - \bi{Y} \bi{F} \right)$, and the
expansion coefficient vector that minimizes the SSR is the familiar
matrix solution of the least squares problem,
\begin{equation}
\bi{F} \,=\, \left( \bi{Y}^\dag \bi{W} \bi{Y} \right)^{-1}
\bi{Y}^\dag \bi{W} \bi{\Phi} \,=\, \bi{G}^{-1} \, \bi{Y}^\dag \bi{W}
\bi{\Phi} \; . \label{Eq:MatrixForm}
\end{equation}
This is obviously equivalent to~(\ref{Eq:AdjointCoefficients})
above.

Let $\bi{P} \equiv \bi{G}^{-1} \, \bi{Y}^\dag \bi{W}$ be the
projection matrix that computes the expansion coefficients when
multiplied by the grid data: $\bi{F} = \bi{P} \,\bi{\Phi}$. As
in~(\ref{Eq:ProjectionFunctions}), appropriate linear combinations
of rows of $\bi{P}$ define a set of row vectors $\bi{p}^{\ell\,m}$
that, when multiplied by the grid data, give the spherical harmonic
amplitudes $\Phi^{\ell\,m}(t,r_0) = \bi{p}^{\ell\,m} \, \bi{\Phi}$.
These projection vectors are computed at the beginning of a
simulation and stored for later use in computing the time-dependent
expansion coefficients as the simulation evolves.

Of course it is not generally advisable to compute $\bi{P}$ by
inverting $\bi{G}$. It is better to compute the $M \times N$ matrix
$\bi{P}$ as the solution of the linear system
\[
\bi{G} \bi{P} = \bi{Y}^\dag \bi{W} \; .
\]
Because $\bi{G}$ is a real, symmetric, positive-definite matrix,
this system can be solved rapidly using Cholesky decomposition.

\subsection{Weighting and Local Regression}

There are actually two types of least-squares regressions involved
in the method: a global regression of all points over the spherical
harmonic basis functions, and a {\it local\/} regression over the
radial basis functions. Local regressions differ from traditional
global regression fits in that they are only used to compute the
best-fit function at a specific point. In this case the local
regression computes the spherical harmonic amplitudes at $r=r_0$
from a fit over the interval $(r_0-\delta,r_0+\delta)$.

Local regression techniques are popular for data smoothing
algorithms~\cite{Cleveland}. In such cases the value of a data point
is replaced by the result of a weighted polynomial fit to it and
neighboring points. A common, though not universal, feature of local
regression techniques is that the weight of each point in the fit
decreases with distance from the evaluation point. Two popular
choices of local weight function are the ``bisquare'' function,
\[
B(u) = \left\{ \begin{array}{l l l}
\left( 1 - u^2 \right)^2 & & | u | < 1 \\
0 & & | u | \geq 1
\end{array} \right. \; ,
\]
and the ``tricube'' function,
\[
T(u) = \left\{ \begin{array}{l l l}
\left( 1 - |u|^3 \right)^3 & & | u_i | < 1 \\
0 & & | u | \geq 1
\end{array} \right. \; ,
\]
where $u = (r-r_0)/\delta$ is a dimensionless radial coordinate
defined so that the domain $(r_0-\delta, r_0+\delta)$ is transformed
to the interval $(-1,1)$.

Contrast this with Misner's method. In the continuum limit, all
points weigh equally in the computation of $\bi{G}$ because the
Legendre polynomials are orthonormal with respect to integration
over a constant weight function.

Now consider Misner's weight in the numerical approximation, written
in terms of the dimensionless radial coordinate $u$. Up to a
multiplicative factor of $h^3$ the weight function is
\[
w(u) = \left\{ \begin{array}{l l l}
0 & & | u | > 1 \\
1 & & | u | < 1 - h/\delta \\
\left( 1 - | u | \right) \delta/h & & \mathrm{otherwise.}
\end{array} \right.
\]
While the weight is constant for points near $r_0$, it falls off
linearly for points near the edges of the shell $S_+$. The degree of
down-weighting depends on the ratio of the size of the shell to the
grid spacing, $\delta/h$. For $\delta \gg h$, the weights are
essentially constant.

\begin{figure}
\begin{center}
\includegraphics[height=3in]{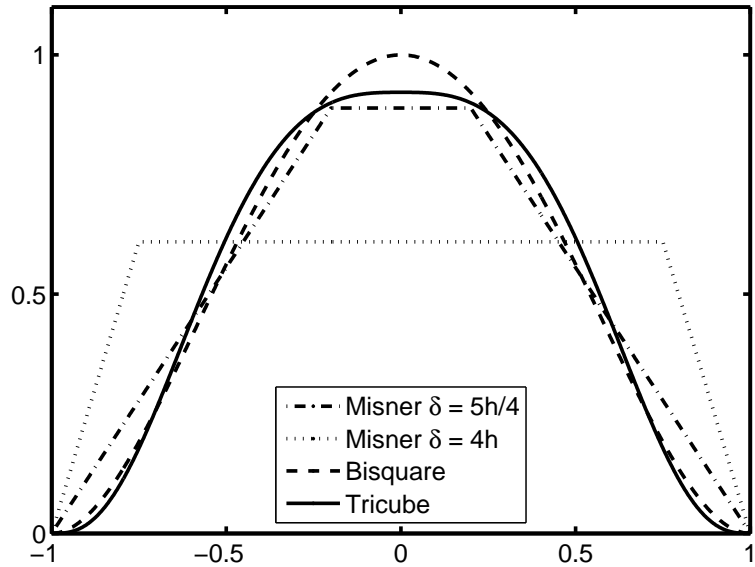}
\end{center}
\caption{\label{Weights}Comparison of Misner's weights and local
weights. The functions are scaled to have the same area underneath
each curve.}
\end{figure}

Figure~\ref{Weights} shows Misner's weight function for two
different values of $\delta / h$ and compared to the bisquare and
tricube weights. In his analysis of Misner's method~\cite{Fiske05},
Fiske finds that $\delta = \frac{5}{4} h$ (equivalently, $\Delta =
\frac{3}{4} h$) gives good results in tests. In this case most
points in $S_+$ are ``near'' the boundary, and the weight function
is local in the sense that it only emphasizes points nearest the
center. In fact, Fiske's choice leads to a weight function that is
very similar in profile to the traditional local regression weights,
and gives similar results.

To see the benefit of using local weights, consider Fiske's analysis
in the continuum limit. Misner's method works because a linear
combination of the radial basis functions $R_n$ form a Dirac delta
function:
\[
\sum_{n=0}^{\infty} \, R_n(r_0) \, R_n(r) = r^{-2} \, \delta(r_0 -
r) \; .
\]
Given the definition of $R_n$ in terms of the Legendre polynomials,
define
\[ d_P(u,N) \equiv \sum_{n=0}^N \left( \frac{2 n + 1}{2} \right)
P_n(0) P_n(u) \; . \] Note that only even values of $n$ figure into
this sum because $P_n(0)=0$ for odd $n$. The function $d_P(u,N)$
will approach the Dirac delta function $\delta(u)$ in the limit as
$N \rightarrow \infty$. The use of a finite number of radial basis
functions produces a truncation error that will have leading term
$\Or(\delta^{N+2})$, so that a fit with $N = 2\mbox{ or }3$ will
converge with shell size as $\Or(\delta^4)$.

The same analysis can be applied using functions that are
orthonormal with respect to integration over the bisquare or tricube
weight functions. Such orthogonal polynomials can be constructed
using a Gram-Schmidt procedure. At least in the case of
bisquare-weight functions, they can also be found by the simple
Rodrigues formula
\[ b_n(u) = c_n \, (1-u^2)^{-2} \, \frac{\rmd^n}{\rmd u^n}
\left( 1 - u^2 \right)^{n+2} \; ,\] where $c_n$ is a normalization
constant.

The first few normalized bisquare-weight polynomials are
\[
\begin{array}{l l}
b_0 (u) =  \sqrt{\frac{15}{16}} &  b_1 (u) = \sqrt{\frac{105}{16}}\,
u \\
b_2 (u) =  \sqrt{\frac{45}{64}} \, \left( 1 - 7 u^2 \right) & b_3
(u) = \sqrt{\frac{1155}{64}} \, \left( u - 3 u^3 \right) \\
b_4 (u)
= \sqrt{\frac{1365}{2048}} \, \left( 1 - 18 u^2 + 33 u^4 \right)\;.
\end{array}
\]
Define
\[ d_b(u,N) \equiv \sum_{n=0}^N b_n(0)\, b_n(u)\, (1-u^2)^2 \; .\]
Again, this will approach $\delta(u)$ in the limit $N \rightarrow
\infty$ and the leading order of truncation error is still
$\Or(\delta^{N+2})$.

\begin{figure}
\begin{center}
\includegraphics[height=3in]{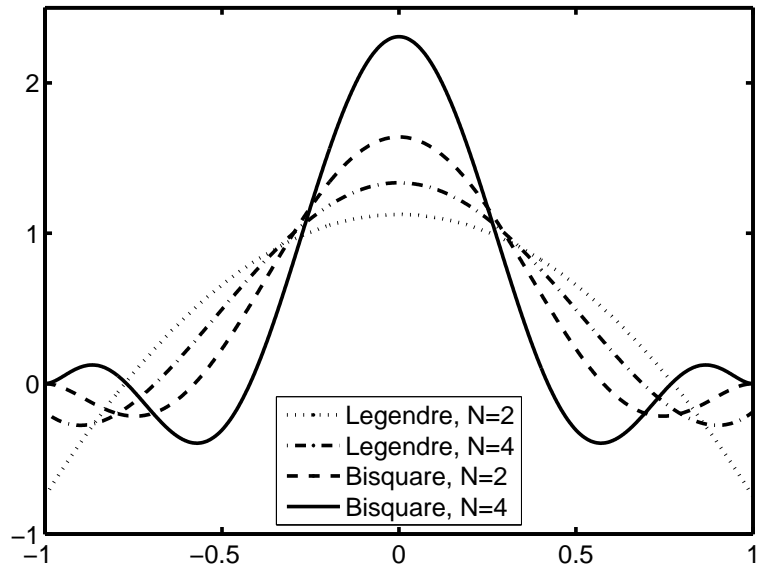}
\end{center}
\caption{\label{DeltaApprox}Comparison of $d(u,N)$ constructed from
Legendre and bisquare-weight polynomials.}
\end{figure}

Figure~\ref{DeltaApprox} compares $d_P$ and $d_b$ for low orders. It
is clear from the figure that the bisquare approximation to
$\delta(u)$ is better at each order than the Legendre approximation.
This is because of the down-weighting inherent in local regressions.

\begin{figure}
\begin{center}
\includegraphics[height=3in]{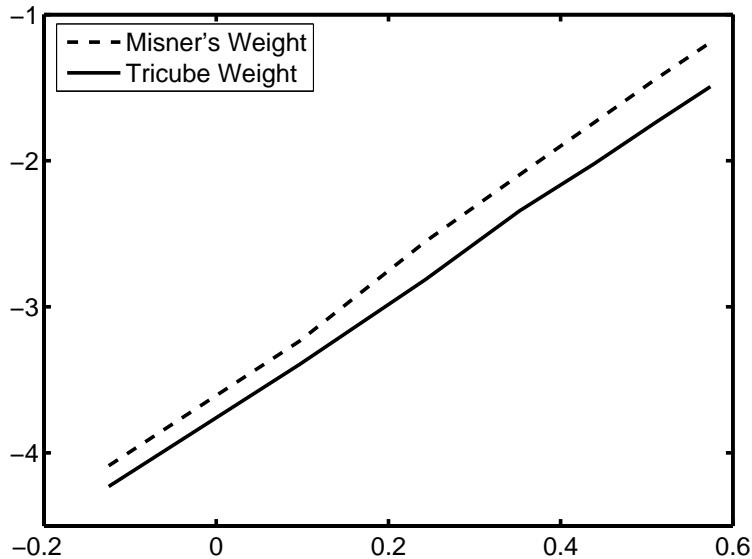}
\end{center}
\caption{\label{Convergedoh}Comparison of $\log(\mathrm{error})$ to
$\log(\delta/h)$ for $h = 1/8$. The tricube error is 4th-order
convergent, and the error for Misner's weight converges slightly
faster.}
\end{figure}

The advantage of the local weighting is apparent in numerical tests.
Figure~\ref{Convergedoh} compares results of computations of
$\Phi^{2 0}$ from the test function using Misner's weight to those
using the tricube weight. (Errors using the bisquare weight are
close to, but slightly larger than, those using the tricube weight.)
The vertical axis is the (base-10) log of the $L_2$ norm of the
error, while the horizontal axis is the log of $\delta/h$. The tests
were performed using a quadratic fit in $r$ ($N=2$).

The error in both plots in figure~\ref{Convergedoh} is fourth-order
convergent with respect to $\delta$ (for fixed $h$), as predicted by
Fiske's analysis. The error using Misner's weight actually converges
slightly faster than fourth-order because the shape of Misner's
weight function changes with $\delta/h$, while the shape of the
tricube weight is fixed. As $\delta$ decreases, Misner's weight
function becomes more local, and it should be no surprise that Fiske
found good results with $\delta = \frac{5}{4} h$. Regardless of the
value of $\delta/h$, however, the bisquare and tricube weights will
give better results than Misner's weight.

\subsection{Switching to the monomial basis}

The preceding analysis using Legendre and bisquare-weight
polynomials is strictly correct in the continuum limit, not for
calculations from grid data. In fact, the properties of Legendre
polynomials in the continuum limit cannot be expected to carry over
into the numerical regime, especially for small values of $\delta$,
because the weight function will no longer be constant. Furthermore,
because Misner's down-weighting prescription for boundary points is
only correct for points near the coordinate axes, the orthonormality
properties of the Legendre polynomials will break down in the
numerical approximation for small shell sizes.

Of course this is not an problem in practice because Misner's method
doesn't really rely on the computation of integrals. Instead, it is
based on solving a weighted least squares problem and the only
requirement for the radial basis functions is that they be linearly
independent. It is quite simple (and useful in practice) to switch
from an orthogonal polynomial basis to a monomial basis:
\begin{equation} R_n(r) =
\left(\frac{r-r_0}{\delta}\right)^n = u^n \;. \label{Eq:Monomial}
\end{equation}
This choice simplifies the calculation by removing the need to form
a linear combination of rows of $\bi{P}$ to get the projection
vectors $\bi{p}^{\ell\,m}$. On the contrary, because $R_0(r_0) = 1$
and $R_n(r_0)=0$ for $n > 1$, then $\Phi^{\ell\,m} =
\Phi^{0\,\ell\,m}$ and the rows of $\bi{P}$ corresponding to $n = 0$
for each $\{\ell,m\}$ are the projection vectors $\bi{p}^{\ell\,m}$.

At this point, one may be concerned that the use of monomial basis
functions might cause $\bi{G}$ to become ill-conditioned. However,
as long as the number of radial basis functions is small, this will
not be a significant problem. Tests have shown that the difference
between using the monomial basis and the orthogonal polynomial basis
for a quadratic fit in $r$ is on the order of the roundoff error
expected from solving a linear system of that size.

An additional benefit of this approach comes in cases where the
radial derivative of $\Phi^{\ell\,m}$ is required at $r_0$. This is
the case, for example, when the extracted multipole amplitudes at
$r_0$ are to be used as inner boundary data for the evolution of a
one-dimensional second-order wave equation for each mode as
in~\cite{POBM}. Based on the definition of $R_n$
in~(\ref{Eq:Monomial}), it is clear that the coefficient
$\Phi^{1\,\ell\,m}$ is the derivative (with respect to $u$) of
$\Phi^{\ell\,m}$. Therefore, we have $\left( \rmd \Phi^{\ell\,m} /
\rmd r \right)_{r=r_0} \, = \, \Phi^{1\,\ell\,m} / \delta$. Again,
this is easily computed using projection vectors that are rows of
$\bi{P}$.

\begin{figure}
\begin{center}
\includegraphics[height=3in]{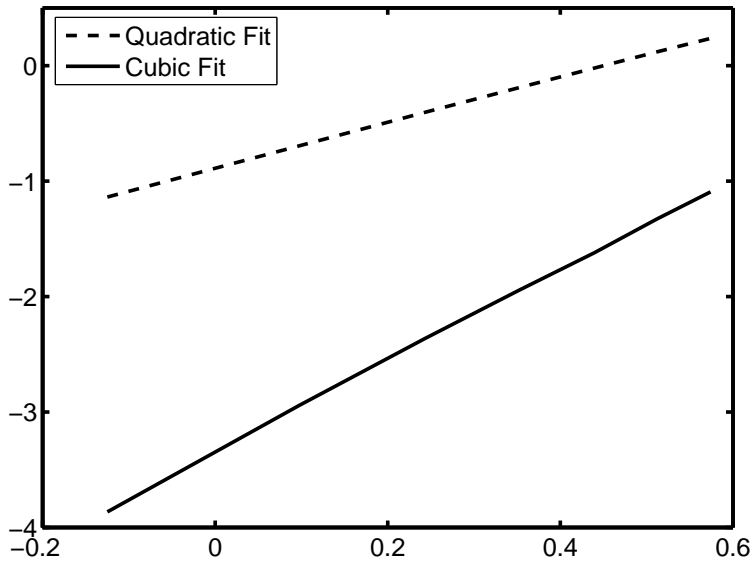}
\end{center}
\caption{\label{Convergederiv}Comparison of errors in computation of
$\rmd \Phi^{2 0} / \rmd r$ for quadratic and cubic fits. The plot
shows $\log(\mathrm{error})$ vs. $\log(\delta/h)$ for $h = 1/8$.}
\end{figure}

Fiske's analysis of truncation error in the approximation to the
Dirac delta function also applies when computing $\rmd
\Phi^{\ell\,m} / \rmd r$, only now it involves odd powers to
approximate the derivative of the delta function. Here the error
will converge as $\Or(\delta^2)$ for a linear or quadratic fit, and
will be $\Or(\delta^4)$ for a cubic fit. Increasing the radial basis
to include $R_3$ will reduce the error in $\Phi^{\ell\,m}$ by only a
very small amount and will not improve the convergence properties of
that error. However using a cubic fit to compute $\rmd
\Phi^{\ell\,m} / \rmd r$ offers significant improvement in both the
error and its convergence properties, as shown in
figure~\ref{Convergederiv}. Here we have a convenient rule of thumb:
a quadratic fit in $r$ is more than sufficient to compute the
multipole amplitudes of $\Phi$, but a cubic fit is strongly
suggested with the derivatives of those amplitudes are also
required.

\subsection{Symmetry issues}

One final advantage of the least-squares perspective comes in its
application to symmetry issues. Fiske~\cite{Fiske05} has studied
extensively the symmetries of the basis functions $Y_A$ and shown
that the dual basis functions, $Y^A$, have the same symmetries. This
is useful in cases where the three-dimensional simulation is
computed in a single octant with explicit reflection symmetries at
the coordinate planes. Fiske's analysis shows how to apply Misner's
method in such cases by exploiting the symmetries of $Y_A$ to
compute the sums used to construct $G_{A B}$, which must be taken
over all octants, using only data defined within the computational
octant. The result is a block diagonal linear system constructed so
that coefficients of harmonics that do not share the octant symmetry
are forced to vanish.

Again, the least squares perspective offers a slightly different
approach to the problem. First, because the least squares fit only
relies on the basis functions being linearly independent, it is
possible simply to perform the fit over all spherical harmonics
using data in the octant. A problem with this approach is that it
could cause $\bi{G}$ to become ill-conditioned as the number of
basis functions grows large.

A simple fix is to restrict the least squares fit to only those
basis functions that share the octant symmetry of the simulation,
and ignore all other basis harmonics. For example, because the test
function in this paper is based on $Y_{2 0}$, the function $\Phi$
will be even under reflection across the coordinate plane boundaries
of the first octant. Using real-valued spherical harmonics, which
are proportional to $\cos(m \phi)$ for positive $m$ and $\sin(m
\phi)$ for negative $m$, it is easy to show that the only harmonic
functions that have this octant symmetry are those with even $\ell$
and even, positive $m$. This guarantees that the fit matrix $\bi{G}$
and the corresponding linear system are much smaller than those
which use all spherical basis functions. A fit with
$\ell_\mathrm{max} = 2$, could potentially involve nine angular
basis functions, but only three of these ($Y_{0 0}$, $Y_{2 0}$, and
$Y_{2 2}$) share the octant symmetry of $Y_{2 0}$. This cuts the
size of linear system by a factor of three, significantly reducing
computation time. While $\bi{G}$ would no longer as sparse as in
Fiske's approach, this should not present a significant additional
computational burden.

\section{Discussion}

Misner's approach to the multipole decomposition of grid data offers
a significant computational advantage over the traditional method of
first interpolating grid data onto the two-sphere, then integrating
against the spherical harmonics. For a shell size $\delta$ that is
proportional to the grid spacing $h$, fourth-order convergence is
achieved by a simple least-squares fit over a quadratic polynomial
in $r$ and the specified set of basis functions (spherical
harmonics) in $\{\theta,\phi\}$. This accuracy is achieved without
the need to consider coordinate patches that cover the two-sphere.

The construction of projection vectors that can be computed at the
beginning of a simulation and stored for use at each computational
time level offers a significant increase in computational speed, as
well. While such operators can in principle be computed for the
interpolation/integration method, their construction is
significantly easier in Misner's method because the projection
operators are simply the result of solving the linear system
associated with a least squares regression problem.

Re-casting Misner's method as a local least squares problem does
offer additional advantages in accuracy and computational
simplicity. Tests suggest that a ($r$-dependent) tricube weighting
scheme gives good results.

This analysis has not investigated the application of Misner's
method to more complex problems that would use, for example, tensor
harmonics. However, because the method is based only on the
orthonormality of the basis functions (indeed, on the weaker
requirement of linear independence) on the two-sphere, it should
extend without difficulty to those problems.

We have also not investigated the application of Misner's method to
mesh refinement. The analysis of Fiske, \etal~\cite{Fiskeetal}
demonstrates the simplicity of Misner's method to cases of {\it
fixed} mesh refinement. Although it is not clear from that paper, a
strict application of Misner's weighting scheme is slightly more
complicated for refined grids than for the fixed grids considered
here. This is because one must account for the fact that points in a
refined region belong to cells with smaller volume $h^3$. However,
from the local least-squares perspective the weight of each point
only depends on its proximity to $r_0$. It is not clear that any
advantage is gained by adjusting the weight of a point based on the
density of grid points nearby.

Application to {\it adaptive} mesh refinement can complicate the
process because there is no way {\it a priori} to establish the
distribution of grid points at any given time level. This would seem
to rule out pre-computing the projection vectors and slow down the
procedure. However, because the computation is done over a shell
that scales quadratically with resolution, this may not represent a
significant time penalty.

More importantly, it may not even be necessary to consider adaptive
mesh refinement when applying Misner's method. As long as the grid
size, number of basis functions, and grid resolution are sufficient
to compute the multipole amplitudes at the most coarse level, the
multipole decomposition can be applied only to grid points on that
level, and these are known {\it a priori}. Unless high-frequency
modes are of interest, the values of the wave on the coarse grid
should be sufficient to compute the multipole amplitudes.

\ack{This research was funded by the FAU Division of Sponsored
Research New Project Development grant \#05-375.}


\Bibliography{9}
\bibitem{POBM}Rupright~M~E, Abrahams~A~M and Rezzolla~L (1998) \PR D
{\bf 58} 044005
\bibitem{Fiskeetal}Fiske~D~R, Baker~J~G, van~Meter~J~R, Choi~D and
Centrella~J~M 2005 \PR D {\bf 71} 104036
\bibitem{LSU}Zink~B, Pazos~E, Diener~P and Tiglio~M 2006
\PR D {\bf 73} 084011
\bibitem{Pitt}G\'omez~R, Lehner~L, Papadopoulos~P and Winicour~J
1997 \CQG {\bf 14} 977--990
\bibitem{Misner04}Misner~C~W 2004 \CQG {\bf 21} S243--8
\bibitem{Fiske05}Fisk~D~R 2005 Error and Symmetry Analysis of
Misner's Algorithm For Spherical Harmonic Decomposition on a Cubic
Grid {\it Preprint} gr-qc/0412047
\bibitem{Cleveland}Cleveland~W~S and Devlin~S~J 1988 {\it J. Amer.
Statist. Assoc.} {\bf 83} 596--610
\endbib
\end{document}